\documentclass{article}
\usepackage{cite}
\usepackage{comment}
\usepackage{amsmath,amssymb,amsfonts}
\usepackage{algorithmic}
\usepackage{graphicx}
\usepackage{textcomp}
\usepackage{xcolor}
\usepackage{arxiv}
\def\BibTeX{{\rm B\kern-.05em{\sc i\kern-.025em b}\kern-.08em
    T\kern-.1667em\lower.7ex\hbox{E}\kern-.125emX}}
\usepackage[symbol]{footmisc}

\title{Empirical study on the efficiency of Spiking Neural Networks with axonal delays, and algorithm-hardware benchmarking}

\author{ Alberto Patiño-Saucedo \\
	CSIC/Universidad de Sevilla\\
	Seville, Spain \\
	\And
        Amirreza Yousefzadeh \\
	IMEC-Netherlands\\
	Eindhoven, Netherlands
 	\And
        Guangzhi Tang \\
	IMEC-Netherlands\\
	Eindhoven, Netherlands\\
 	\And
        Federico Corradi \\
	Eindhoven University of Technology\\
	IMEC-Netherlands\\
 	\And
        Bernabé Linares-Barranco \\
	CSIC/Universidad de Sevilla\\
	Seville, Spain \\
 	\And
        Manolis Sifalakis \\
	IMEC-Netherlands\\
	Eindhoven, Netherlands\\
}

\begin{document}

\maketitle

\begin{abstract}

The role of axonal synaptic delays in the efficacy and performance of artificial neural networks has been largely unexplored. In step-based analog-valued neural network models (ANNs), the concept is almost absent. In their spiking neuroscience-inspired counterparts, there is hardly a systematic account of their effects on model performance in terms of accuracy and number of synaptic operations.This paper proposes a methodology for accounting for axonal delays in the training loop of deep Spiking Neural Networks (SNNs), intending to efficiently solve machine learning tasks on data with rich temporal dependencies. We then conduct an empirical study of the effects of axonal delays on model performance during inference for the Adding task~\cite{HockreiterSchmidhuber1997,Bai2018,kag2021training}, a benchmark for sequential regression, and for the Spiking Heidelberg Digits dataset (SHD)~\cite{zenke2022shd}, commonly used for evaluating event-driven models. Quantitative results on the SHD show that SNNs incorporating axonal delays instead of explicit recurrent synapses achieve state-of-the-art, over 90\% test accuracy while needing less than half trainable synapses. Additionally, we estimate the required memory in terms of total parameters and energy consumption of accomodating such delay-trained models on a modern neuromorphic accelerator~\cite{SENeCA,tang2023seneca}. These estimations are based on the number of synaptic operations and the reference GF-22nm FDX CMOS technology. As a result, we demonstrate that a reduced parameterization, which incorporates axonal delays, leads to approximately 90\% energy and memory reduction in digital hardware implementations for a similar performance in the aforementioned task.
\end{abstract}





\section{Introduction}

Spiking Neural Networks (SNNs) are models more closely resembling biology than Analog Neural Networks (ANNs) due to their statefulness and binary event-driven encoding of information, which on novel neuromorphic processors, render them highly efficient in temporal processing applications. 
Lending to more (compact) parameterization SNNs demonstrate competitive performance to deep ANNs (DNNs)~\cite{yin2021accurate}; while potentially using fewer MAC operations in digital hardware implementations.
Furthermore, the statefulness of SNNs, embodied in the (decaying) membrane potential of neurons, allows them to be mapped to RNNs~\cite{neftci2019surrogate} effectively, even without recurrent synaptic connections. However, for temporal tasks, the best-performing SNN models almost universally include explicit recurrent connections \cite{yin2021accurate, yin2020effective, zenke2021remarkable, zenke2022shd, perez2021neural}, which exponentially increases the number of required synaptic weights as a function of the number of neurons, adding a burden to neuromorphic hardware development.

Meanwhile, the role of axonal delays, i.e., the delay  for a spike (action potential) to travel from the soma to the axon terminals, which is a critical element of parameterization in biological neural networks, has remained largely unexplored or characterized in the study of the efficacy, model size, and performance of SNNs. This paper attempts an initial characterization of and effects of synaptic delays on SNN model performance and the impact of accounting for them in neuromorphic processor architectures.

The first contribution of the work in this paper is a simple strategy of training SNN models with axonal delays, which is conformal with back-propagation (BP) frameworks commonly used for SNN/DNN training (BP through-time (BPTT) for DNNs and its extension spatio-temporal BP (STBP) for SNNs). The second contribution regards an assessment and quantification of the effects of  synaptic delay parameterization on model performance (accuracy), model complexity (network structure) and model size (number of parameters). The third contribution is a quantification of energy and memory cost of deploying models with synaptic delays on a modern neuromorphic processor, based on two different design strategies.



\section{Related Work}

Perhaps one of the reasons that delay model training has not been as mainstream in artificial neural network research until now, is the fact that ANN accelerators do not specifically account and optimize for them at the hardware level. By contrast, many digital neuromorphic accelerators provide explicit hardware support for delay structures (dendritic/axonal); either per neuron~\cite{SpiNNaker, Loihi, morrison2005advancing}, or shared across neurons~\cite{TrueNorth, SENeCA}. This makes delay model training an attractive exploration in relation to compute and power efficiency.

Recurrency in neural networks offers a constrained way of compensating for synaptic delay parameterization, limited to a single-timestep. Despite this limitation, only a handful of works have explored the explicit use of synaptic delays independently of recurrences. One common formalization in the literature of TDNNs \cite{Day1993,DuroReyes1999,Santos2017} and delay-aware SNNs \cite{Shrestha2018,Wang2019,Schrauwen2004,Shrestha2016} is to parameterize synapses with an additional learnable delay variable, trainable with back-propagation \cite{BoneCardot2005, Wang2019}, local Hebbian-like learning rules \cite{Zhang2020}, or annealing algorithms \cite{Cohen1997}. An alternative approach in TDNNs involves mapping delays in the spatial domain and train them with autoregressive models and so-called temporal convolutions (TCNs) \cite{Waibel1989,Vandenoord2016,Gregor14,Bai2018,Lea2017,Peddinti2015}. This approach enables structurally simpler models, which are easier/faster to train, but not very compact as their breadth/depth must scale linearly with the number of timesteps needed to capture temporal dependencies. Our approach is akin to this latter strategy but because of the incremental delay quantization-pruning, our models neither narrow the aperture of the temporal window nor make it homogeneous for all neurons (does not lead to deep models). 


\section{Methods}

\subsection{Delay Model Description}

We use multilayer Leaky Integrate-and-Fire (LIF) Spiking Neural Networks (SNNs). LIF neurons are stateful, and represent a compromise between biological plausibility and computational efficiency for hardware implementation. Their excitation depends on both their time-dependent input $I$ from other neurons and on their internal state, known as the membrane potential $u$ subject to leaky integration with a time constant ${\tau}$. The equations of the membrane potential update in a discrete-time implementation of a LIF spiking neuron are:

\begin{equation}
    u_k = u_{k-1}e^{-\frac{1}{\tau}} (1-\theta_{k-1}) + I_{k-1}
\end{equation}

\begin{equation}
\theta_{k} = \begin{cases}
1 & {u}_{k} \geq u_{th} \\ 
0 & \text{otherwise}
\end{cases}
\end{equation}

\noindent where ${\theta}$ denotes a function to generate activations or spikes whenever the membrane potential reaches a threshold associated with the neuron, $u_{th}$. 

\begin{figure}
    \centering
    \includegraphics[width=12cm]{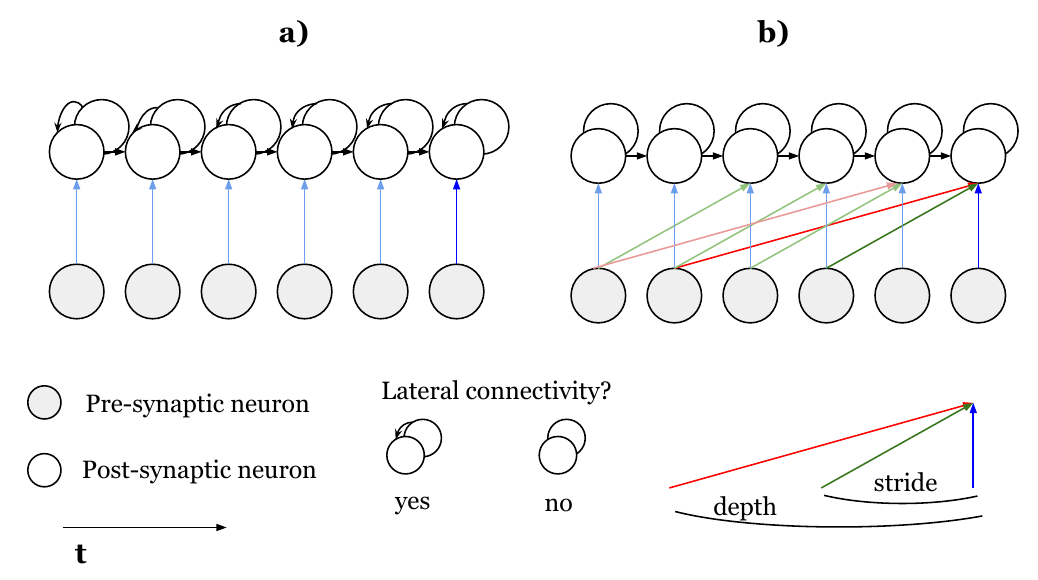}
    \caption{Projections over time in a pre/post-synaptic pair for a) Recurrently connected SNN (R-SNN ) and b) Delayed SNN (D-SNN) using receptive fields of stride 2 and depth 4. Weight values are color-consistent.}
    \label{fig:Comp_DSNN_RSNN}
\end{figure}

Multilayer SNNs can be organized as feedforward or recurrent networks. In a recurrent SNN layer, neurons exhibit lateral connectivity, as in Fig.~\ref{fig:Comp_DSNN_RSNN}(a), and their input is computed by adding the weighted contribution from the $N$ neurons to the previous or pre-synaptic layer and from the $M$ neighboring neurons in their own layer, as shown in the next equation:

\begin{equation}
I_k{[recurrent]} = \sum_{i=1}^{N} w_{i} \theta_{i,k} + \sum_{j=1}^{M} w_{j} \theta_{j,k}
\end{equation}

To incorporate axonal delays in networks of spiking neurons, we create multiple time-delayed projections or synapses for every pre-synaptic/post-synaptic neuron pair. This way, the activation of a neuron at a given time depends on both its current state and a subset of past activations from neurons in the pre-synaptic layer, with direct projections. The input of a neuron incorporating the proposed model for axonal delays is: 

\begin{equation}
I_k{[delays]}= \sum_{d \in D} \sum_{i=1}^{N} w_{i, d} \theta_{i,k-d}
\end{equation}

\noindent where  $D \in  [0,T]$ is the set of delays chosen for a given task. Control over the temporal stride of the delays and the depth of the temporal receptive field is included in the model (see Fig.\ref{fig:Comp_DSNN_RSNN}(b) for a visualization of the concept). This increases flexibility in the total number of parameters.



\subsection{Model Training}

We train models that incorporate axonal delays using the following approach, which is compatible with vanilla back-propagation frameworks used to train SNNs and RNNs today (i.e. no special framework extensions).
The idea is to express the (temporal) parameterization of delays as a spatial parameterization of synaptic weights, such that delay training is effected by merely optimizing for synaptic weights. We start with a set of parallel synapses per pair of pre and post-synaptic neurons each associated with a delayed output from the pre-synaptic neuron (using a predetermined range of delays and stride). We optimize the model as usual, and prune all delay-synapses that end up with small weights. We then fine-tune the model with only the remaining synapses. We may introduce new synapses to replace the pruned ones, with incrementally higher delay resolution in localized sub-regions of the initial delay range, and repeat the process. As a result different neurons end-up with different fan-in delay inputs. The resulting models are topologically feed-forward, consistently shallower with few parameters than their recurrent-connectivity counterparts, and exhibit state-of-art performance (confirmed in all experiments). Their simpler structure renders them attractive for resource-efficient deployment on neuromorphic accelerators.

We trained SNN models with back-propagation (STBP specifically~\cite{wu2018stbp}). This method accounts for the past influence on current neurons' states by unrolling the network in time, and the errors are computed along the reverse paths of the unrolled network. To account for the discontinuity of the membrane potential, we employed a surrogate gradient function~\cite{neftci2019surrogate} with a fast sigmoid function as in ~\cite{zenke2021remarkable}. During training, apart from the synaptic weights, we also optimized the membrane's time constants, as in~\cite{yin2021accurate}. Finally, we did not consider extra delays for the input at the first layer, as the input layer usually has more neurons and is responsible for a large portion of the synaptic parameters.

\subsection{Implementation cost in neuromorphic processors}
\label{sec:delayimplem}

\begin{figure}
    \centering
    \includegraphics[width=12cm]{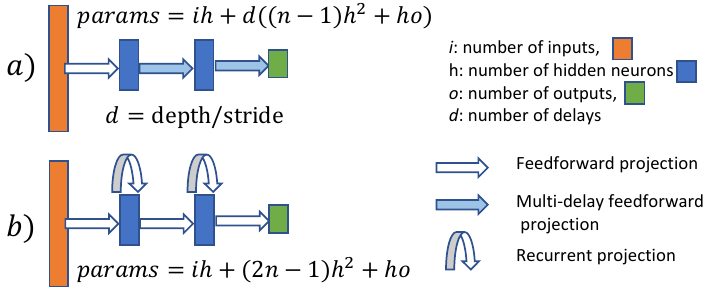}
    \caption{Equations used to calculate the max number of parameters for (a) the delay-based architecture and (b) recurrent SNN proposed here.}
    \label{fig:param_formula}
\end{figure}

Neuromorphic hardware architectures implement stateful nodes with scalable event-driven communication, reducing communication and processing costs and, by extension, the required energy. Spiking Neural Networks are some of the most well-suited algorithms for these kinds of processors, and as such, the delay mechanism is supported by most neuromorphic chips. 
In this paper, we used a simple yet accurate methodology to compare the energy and memory overhead of the delay mechanism. The energy consumption is calculated based on counting the memory accesses (spike packets and neuron states read/write and weights read) and arithmetic operations (accumulation, comparison with threshold, etc.) using a netlist level simulation tool (Cadence JOULES) for an advanced technology node (Global-Foundries 22nm FDX). Memory cost is calculated from the total number of parameters (as shown in Fig.\ref{fig:param_formula}), the neuron states, and the number of delayed spike packets required to perform inference. We explored two methods commonly used by digital neuromorphic platforms to implement delay: The Ring Buffer~\cite{SpiNNaker, Loihi, morrison2005advancing} and the Delay Queue ~\cite{TrueNorth,SENeCA}.

\subsubsection{Ring Buffer}

\begin{figure}
    \centering
    \includegraphics[width=12cm]{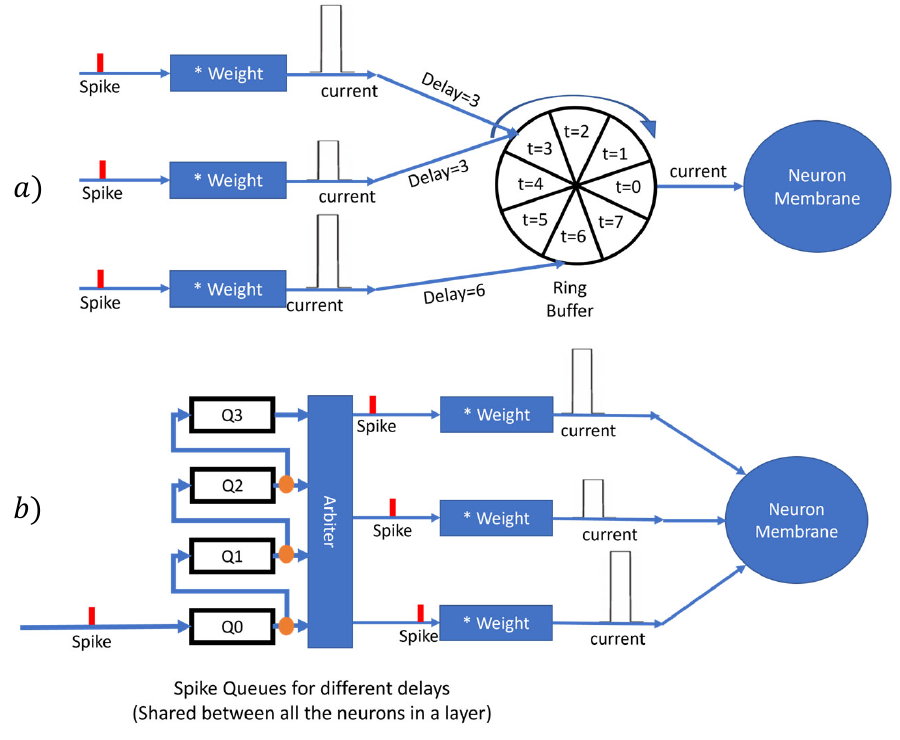}
    \caption{(a) Using a ring buffer per neuron to implement synaptic delays. Unit of delay is the system time-step. (b) Implementation of axon delay by sorting spikes based on the encoded delays in separated delay queues. The queues are shared across neurons in a neuro-synaptic core (not shown in figure).}
    \label{fig:delay_queue_buffer}
\end{figure}

A ring buffer is a special type of circular queue where currents with different delays accumulate in separate elements of the queue. When using the ring buffer, the maximum possible delay in the system will be limited to the size of the buffer, and the set of possible delays is linearly distributed i.e., the temporal stride is constant (see Fig.\ref{fig:delay_queue_buffer}(a)). In this method, there is one ring buffer per neuron; therefore, the memory overhead scales with the number of neurons. 

The estimated memory overhead for the ring buffer (total sum of the ring buffer sizes) is calculated as ``number of postsynaptic neurons with synaptic delay $\times$ maximum synaptic delay". The energy overhead is equal to one extra neural accumulation per time step (to accumulate the value of the ring buffer into the membrane potential). 

\subsubsection{Delay Queue}

The axon delay is encoded directly in the spikes in a delay queue. Therefore, each spike packet contains a few bits to indicate the amount of delay. In the destination neuro-synaptic core, instead of having a single queue for all spikes, several queues, each corresponding to a specific delay amount, are implemented. This method is more efficient to implement when spikes activity is sparse. Fig\ref{fig:delay_queue_buffer}(b) depicts an implementation of four delay queues. These delay queues are cascaded, are shared by many neurons, and encode an arbitrary amount of delay (does not need to be a linear distribution). In this scheme, unlike the ring buffer, the number of queues is defined based on the number of possible delays and not on the maximum delay amount. However, the size of each queue increases if the queue applies more delay on the spikes (which means the queue needs to keep the spikes for a longer period). Additionally, this method implements the axon delay which is more coarse-grained compared to the dendritic delay implemented by the ring buffer.


To calculate the memory overhead of delay queues, we need to know the number and size of each queue. We assumed that the delay queues are shared between the neurons of a layer. The number of queues is equal to the number of possible delays. Also, since the proposed algorithm assumes that all input spikes are delayed evenly, the total size of all delay queues is equal to the ``maximum number of input spikes of the layer in all time-steps $\times$ the maximum amount of delay". In this way, there is enough space in the queue to keep the delayed spikes for each time step. We estimate the energy overhead from total number of reads and writes to the delay queues.  
 
\section{Results}

We report experiments that demonstrate qualitatively the advantages of training SNN models with axonal delays, and quantitatively the benefits from deploying them in digital neuromorphic processors. The first experiment illustrates that models with axonal delays encode more effectively long-term dependencies than networks with recurrent connections. The second experiment reveals that models with synaptic delays achieve state-of-the-art performance, in tasks rich with temporal features, while requiring fewer parameters than recurrent models (similar observations were confirmed on other datasets). This alludes to more compact models, that require less resources for executing on hardware accelerators. A third experiment quantifies this intuition by means with estimates of energy and memory cost, showing a reduction by an order of magnitude, when such models are employed on neuromorphic accelerators, by comparison to \emph{equi-performing} models with recurrent connections. All models were training with the deep learning framework PyTorch on Nvidia GeForce RTX GPU.

\subsection{Adding task}

\begin{figure}
    \centering
    \includegraphics[width=12cm]{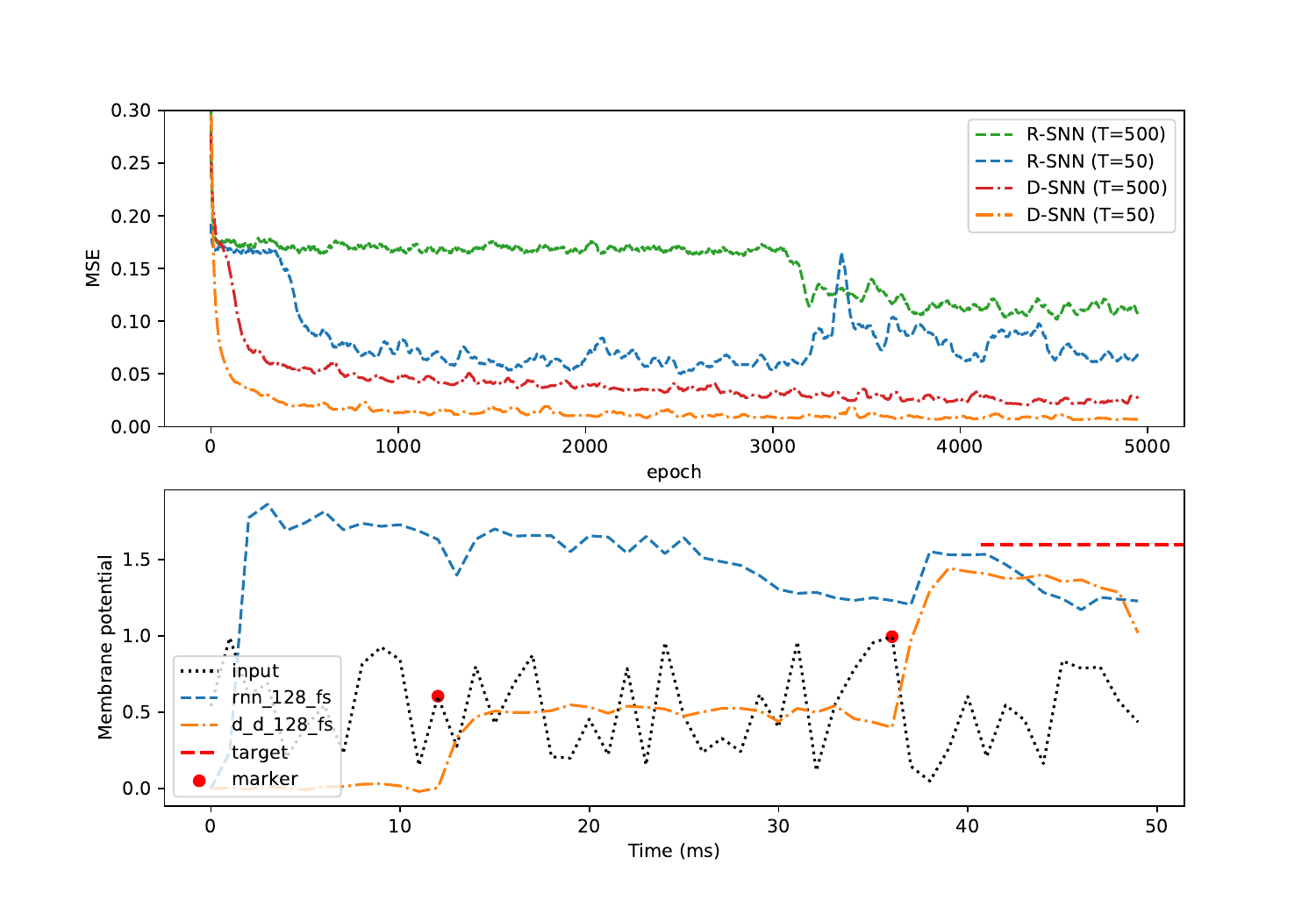}
    \caption{Top: An example of the Adding Task for a sequence length T=50, solved by an R-SNNN (orange) and a D-SNN (green). Notice how the D-SNN "remembers" both values relevant to the task in a more natural way. Bottom: MSE per training epoch for R-SNN (with recurrent synapses) and D-SNN (with delay synapses) in the Adding task, for two sequence lengths: T=50 and T=500. D-SNNs converge faster and to a smaller error!}
    \label{fig:add_mems}
\end{figure}

The~\textit{adding task} is a known benchmark used to evaluate the performance of neural network models on sequence data, such as LSTM~\cite{HockreiterSchmidhuber1997} or TCN~\cite{Bai2018,Lea2017}. The input data is a stream of random values chosen uniformly in [0,1], and two randomly selected indexes, one for every half of the sequence. The target, which should be computed at the end of the sequence, is the addition of the two values of the stream at the selected indexes. To use this task to evaluate generic SNNs, we feed the network through two input channels, one for the number stream and the other for the binary-encoded markers, and then compute the Mean Squared Error (MSE) between the target and the membrane potential of a readout neuron with an infinite threshold.
%
%
Fig. \ref{fig:add_mems} (top) shows that while both a recurrent connectivity and a delay-synapses enable an SNN to remember the indexed numbers and compute the result, the latter however exhibits a more ``sensible'' or interpretable evolution towards the answer. The bottom of the figure on the other hand, reveals that models with synaptic delays converge typically much faster than traditional ones with recurrent connectivity.



\subsection{SHD task}

\begin{figure}
    \centering
    \includegraphics[width=12cm]{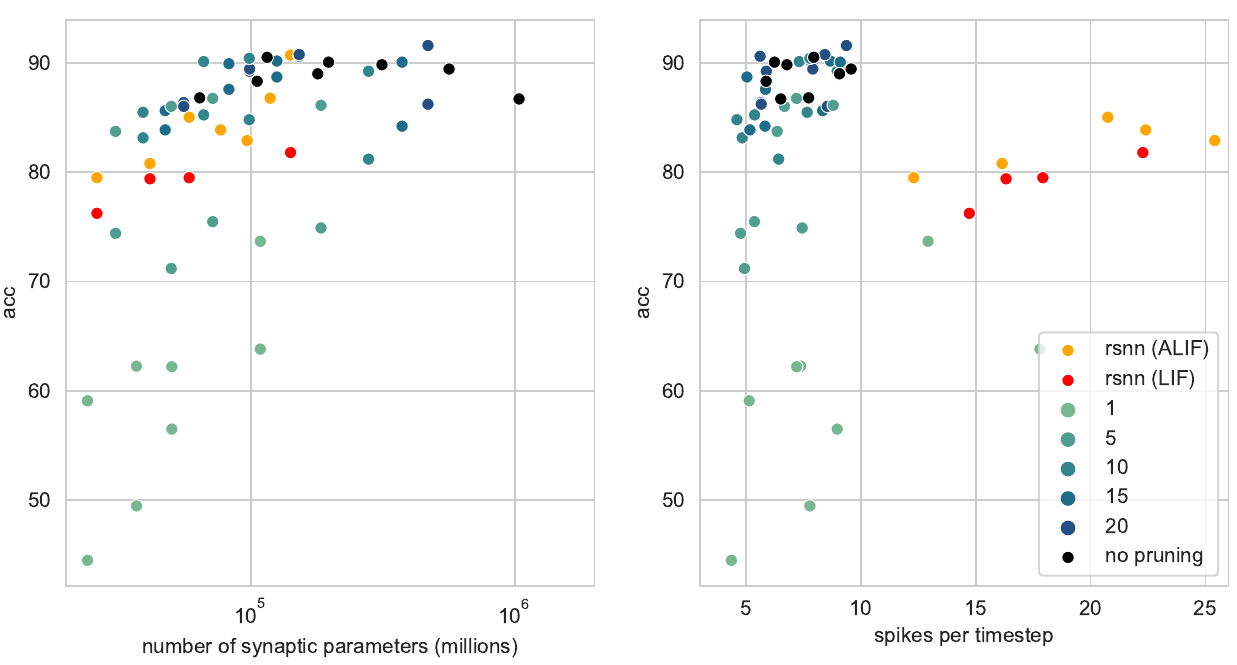}
    \caption{Effect of synaptic delays on performance (SHD task). Left: num of parameters vs accuracy. Right: num of spikes vs accuracy. Red and orange points are recurrently connected SNNs. Colors ranging from green to black are SNNs with axonal delays and different pruning configurations.}
    \label{fig:shd_results}
\end{figure}

Fig.~\ref{fig:shd_results} shows for different models, a comparison of the accuracy on the SHD dataset~\cite{zenke2022shd} as a function respectively of the number of model parameters and the number of spikes generated by the model at inference (the number of parameters is a proxy metric for the model size/complexity, and spikes is a proxy metric of energy consumption on any hardware accelerator). The comparison includes various models generated with our method while bounding the max numbers of delay synapses per neuron pair retained after pruning. No pruning refers to retaining all delay synapses. The comparison includes as baseline two recurrent SoA models from the literature~\cite{yin2021accurate} that use the adaptive LIF (ALIF) and LIF neuron models. The observation is that with the herein proposed training method we can generate models, which are exceptionally compact, energy-efficient, and yet achieve SoA accuracy. These results are further quantified and distilled in Table~\ref{tb:shd_results}, where a comparison is made with different feed-forward and recurrent SNN architectures found in the literature for the same dataset.

\subsection{Energy estimations of hardware implementation}

Table~\ref{tb:energy_results} reports the proposed algorithm's estimated energy consumption and memory footprint for both of the commonplace implementations of delay synapses in existing neuromorphic processors discussed in section~\ref{sec:delayimplem}. The main take-away observation is that the energy and memory overhead from utilizing synaptic delay hardware structures is substantially off-set by the far more compact, with sparser activity, synaptic delay models. The energy estimations are provided only for comparison purposes and extracted from simulations of digital circuits (SRAM memory accesses and arithmetic operations in float 16b data type). For memory overhead, we assumed that all parameters, neuron states, and spike packets use the same data types and only report the total number of memory words. Simulations are for CMOS digital technology node GF-22nm FDX, through Cadence software tools.

\begin{table}[]
\caption{Comparing accuracy and number of parameters for SHD.}
\label{tb:shd_results}
\begin{center}
\begin{tabular}{|l|l|c|r|r|r|}
\hline
\textbf{Paper} & \textbf{Neuron   Type} & \multicolumn{1}{c|}{\textbf{Architecture$^{\ast}$}} & \textbf{T} & \textbf{Params.} & \textbf{Acc.}  \\ \hline
Eshraghian, 2022   & LIF$^{\mathrm{a}}$            & 3000r                                      & 100        & 11160000         & 83.2           \\ \hline
Eshraghian, 2022 & LIF$^{\mathrm{a}}$                & 3000                                       & 100        & 2160000          & 66.3          \\ \hline
Bauer, 2022 & SRM                    & 100+1281+1281+201                          & 250        & 2100562          & 78.1           \\ \hline
Zenke, 2022 & LIF                   & 1024r                                      & 2000       & 1785856          & 83.2           \\ \hline
Fang, 2021 & SRM                  & 400+400                                    & 2000       & 448000           & 85.7          \\ \hline
Yu, 2022 & LIF$^{\mathrm{b}}$       & 400+400                                    & 1000       & 448000           & 87.0             \\ \hline
Zenke, 2021 & LIF                  & 256r+256r                                  & 500        & 380928           & 82.0             \\ \hline
Yin, 2020 & LIF$^{\mathrm{c}}$                & 256r                                       & 250        & 249856           & 81.7          \\ \hline
Yin, 2021  & LIF$^{\mathrm{c}}$               & 128r+128r                                  & 250        & 141312           & \textbf{90.7}  \\ \hline
Zenke, 2022 & LIF                    & 128r                                       & 2000       & 108544           & 71.4           \\ \hline
Perez, 2021 & LIF                    & 128r                                       & 2000       & 108544           & 82.7           \\ \hline
Ours (1) & \textbf{LIF}             & 64d+64d                                    & 250        & 98560        & \textbf{90.4} \\ \hline
Ours (2) & \textbf{LIF}                      & 48d+48d                                    & 250        & \textbf{66240}   & 90.1          \\ 
\hline
\multicolumn{4}{l}{$^{\ast}$ Conventions: r: with lateral recurrency, d: with delay synapses.}  \\
\multicolumn{4}{l}{$^{\mathrm{a}}$ Binarized. $^{\mathrm{b}}$ MAP-SNN. $^{\mathrm{c}}$ Adaptive threshold. }  \\
\end{tabular}
\end{center}
\end{table}

\begin{table}[]
\caption{Energy and memory estimations for the proposed network, compared to an RSNN for similar accuracy.} 
\label{tb:energy_results}
\begin{center}
\begin{tabular}{|lcccc|}
\hline
\multicolumn{1}{|l|}{\textbf{Measurement}}        & \multicolumn{1}{c|}{\textbf{R1}} & \multicolumn{1}{c|}{\textbf{R2}} & \multicolumn{1}{c|}{\textbf{D1}} & \multicolumn{1}{c|}{\textbf{D2}} \\ \hline
\multicolumn{1}{|l|}{neurons per   hidden layer}  & \multicolumn{1}{c|}{128}         & \multicolumn{1}{c|}{48}          & \multicolumn{1}{c|}{8}           & 8                                \\ \hline
\multicolumn{1}{|l|}{number of   delays}          & \multicolumn{1}{c|}{1}           & \multicolumn{1}{c|}{1}           & \multicolumn{1}{c|}{10}          & 5                                \\ \hline
\multicolumn{1}{|l|}{avg   spk/timestep, layer1}  & \multicolumn{1}{c|}{8.678}       & \multicolumn{1}{c|}{6.725}       & \multicolumn{1}{c|}{1.894}       & 1.686                            \\ \hline
\multicolumn{1}{|l|}{avg   spk/timestep, layer 2} & \multicolumn{1}{c|}{4.582}       & \multicolumn{1}{c|}{3.456}       & \multicolumn{1}{c|}{1.772}       & 2.539                            \\ \hline
\multicolumn{1}{|l|}{max   spk/timestep, layer 1} & \multicolumn{1}{c|}{-}           & \multicolumn{1}{c|}{-}           & \multicolumn{1}{c|}{7}           & 7                                \\ \hline
\multicolumn{1}{|l|}{max   spk/timestep, layer 2} & \multicolumn{1}{c|}{-}           & \multicolumn{1}{c|}{-}           & \multicolumn{1}{c|}{7}           & 8                                \\ \hline
\multicolumn{1}{|l|}{test set   accuracy}         & \multicolumn{1}{c|}{81.020}      & \multicolumn{1}{c|}{80.200}      & \multicolumn{1}{c|}{82.170}      & 80.510                           \\ \hline
\multicolumn{5}{|c|}{\textbf{Neurosynaptic cost estimation}}                                                                                                                                  \\ \hline
\multicolumn{1}{|l|}{energy (uJ)}                 & \multicolumn{1}{c|}{20.213}      & \multicolumn{1}{c|}{7.390}       & \multicolumn{1}{c|}{2.304}       & 1.745                            \\ \hline
\multicolumn{1}{|l|}{memory (param.   count)}     & \multicolumn{1}{c|}{141588}      & \multicolumn{1}{c|}{41684}       & \multicolumn{1}{c|}{7876}        & 6756                             \\ \hline
\multicolumn{5}{|c|}{\textbf{Delay queue   estimations}}                                                                                                                                      \\ \hline
\multicolumn{1}{|l|}{energy   overhead (uJ)$^{\ast}$}      & \multicolumn{1}{c|}{\textbf{-}}  & \multicolumn{1}{c|}{\textbf{-}}  & \multicolumn{1}{c|}{0.059}       & 0.030                            \\ \hline
\multicolumn{1}{|l|}{mem. overhead   (words)}     & \multicolumn{1}{c|}{-}           & \multicolumn{1}{c|}{-}           & \multicolumn{1}{c|}{1890}        & 1800                             \\ \hline
\multicolumn{1}{|l|}{energy saving   factor}      & \multicolumn{1}{c|}{1}           & \multicolumn{1}{c|}{2.735}       & \multicolumn{1}{c|}{8.554}       & 11.384                           \\ \hline
\multicolumn{1}{|l|}{memory saving   factor}      & \multicolumn{1}{c|}{1}           & \multicolumn{1}{c|}{3.397}       & \multicolumn{1}{c|}{14.498}      & 16.548                           \\ \hline
\multicolumn{5}{|c|}{\textbf{Ring buffer   estimations}}                                                                                                                                      \\ \hline
\multicolumn{1}{|l|}{energy   overhead (uJ)}      & \multicolumn{1}{c|}{-}           & \multicolumn{1}{c|}{-}           & \multicolumn{1}{c|}{0.085}       & 0.085                            \\ \hline
\multicolumn{1}{|l|}{mem. overhead   (words)}     & \multicolumn{1}{c|}{-}           & \multicolumn{1}{c|}{-}           & \multicolumn{1}{c|}{3780}        & 3360                             \\ \hline
\multicolumn{1}{|l|}{energy saving   factor}      & \multicolumn{1}{c|}{1}           & \multicolumn{1}{c|}{2.735}       & \multicolumn{1}{c|}{8.463}       & 11.046                           \\ \hline
\multicolumn{1}{|l|}{memory saving   factor}      & \multicolumn{1}{c|}{1}           & \multicolumn{1}{c|}{3.397}       & \multicolumn{1}{c|}{12.147}      & 13.996                           \\ \hline
\multicolumn{5}{l}{$^{\ast}$The energy overhead is calculated per inference.} \\
\multicolumn{5}{l}{All networks evaluated for T=250. Columns:} \\
\multicolumn{5}{l}{R1: (Recurrent) LIF 128r+128r.} \\
\multicolumn{5}{l}{R2: (Recurrent) ALIF 48r+48r. } \\
\multicolumn{5}{l}{D1: (Delays) LIF 8d+8d, depth=150, stride=15.} \\
\multicolumn{5}{l}{D2: (Delays) LIF 8d+8d, depth=150, stride=30.}
\end{tabular}
\end{center}
\end{table}

\section{Conclusion}

We introduced a method for training SNN models with synaptic delays, and we report benefits of deploying such models in neuromorphic accelerators.
The important observation from the resulting trained models is that even a small set of synaptic delays together with trainable time constants, supersede the need for complex lateral connectivity, reduce the number of layers and total number of parameters needed for good performance. This also reduces the memory footprint of these models in neuromorphic accelerators (compared to commonplace RNNs).
Future work will focus on \emph{hardware-aware} training of synaptic delay models for compact mappings on neuromorphic accelerators.

\bibliographystyle{IEEEtran}
\bibliography{main}

\footnote[1]{This work was funded by EU grants 824164, 871371, 871501, 89955.}

\end{document}